\documentclass[aps,prl,twocolumn,showpacs,superscriptaddress,groupedaddress]{revtex4-1}  
\usepackage{dcolumn}   
\usepackage{bm}        
\usepackage{amssymb}   
 \usepackage{amsmath}
\usepackage{exscale}
\usepackage{bm}
\usepackage{natbib}
\usepackage{float}
\usepackage[dvips,final]{graphicx}
\usepackage{graphicx}
\usepackage[tight,TABTOPCAP]{subfigure}
\usepackage{pslatex}
\usepackage{relsize}
\usepackage{bm}
\usepackage{xspace} 

\usepackage{xcolor}
\usepackage{soul}

\def\ba{\begin{eqnarray}}
\def\ea{\end{eqnarray}}
\def\be{\begin{equation}}
\def\ee{\end{equation}}

\bibpunct{[}{]}{,}{n}{}{} 

\begin{document}

\title{Robust dynamics of antiferromagnetic skyrmion driven by spin-polarized current in small thin disks}

 \date{\today}

\author{R. L. Silva} \email{ricardo.l.silva@ufes.br}
\affiliation{Departamento de Ci\^{e}ncias Naturais, Universidade Federal do Esp\'{i}rito Santo, Rodovia Governador M\'{a}rio Covas, Km 60, S\~{a}o Mateus, ES, 29932-540, Brazil.}

\author{R. C. Silva} \email{rodrigo.c.silva@ufes.br}
\affiliation{Departamento de Ci\^{e}ncias Naturais, Universidade Federal do Esp\'{i}rito Santo, Rodovia Governador M\'{a}rio Covas, Km 60, S\~{a}o Mateus, ES, 29932-540, Brazil.}

\author{A. R. Pereira} \email{apereira@ufv.br}
\affiliation{Universidade Federal de Vi\c{c}osa, Departamento de F\'{i}sica, Avenida Peter Henry Rolfs s/n, 36570-000, Vi\c{c}osa, MG, Brazil}

\author{W. A. Moura-Melo} \email{w.mouramelo@ufv.br}
\affiliation{Universidade Federal de Vi\c{c}osa, Departamento de F\'{i}sica, Avenida Peter Henry Rolfs s/n, 36570-000, Vi\c{c}osa, MG, Brazil}

\begin{abstract}
We investigate skyrmion configuration and dynamics in antiferromagnetic thin disks. It is shown that the skyrmion acquires oscillatory dynamics with well-defined amplitude and frequency which may be controlled on demand by the spin-polarized current. Such dynamics are robust in the sense that an interface between two half-disks cannot change the dynamics appreciably. Indeed, the skyrmion keeps its oscillatory despite crossing this interface. The way skyrmion found to do that is by modifying its core region shape so that its total energy is unaltered for several cycles.
\end{abstract}

\maketitle

\section{Introduction and Motivation}

The study of topological excitations is an important topic in modern theoretical and experimental physics. It can be relevant for technological applications in several branches of condensed matter physics such as superconductivity, superfluidity, magnetism, etc. They frequently arise in the form of domain-walls, vortices, and solitons in magnetic systems. For instance, solitons (nowadays frequently termed as skyrmions) were described by Belavin and Polyakov\cite{Belavin} within the ferromagnetic Heisenberg framework as topologically protected spin textures with a quantized topological number\cite{Nagaosa1}, even though earlier independent studies by Feldtkeller and Thiele investigated similar patterns \cite{Angew, Thiele}. Magnetic skyrmions are whirl-like quasiparticles on the sub-micrometer scale, first observed in $MnSi$ \cite{Muhlbauer} as periodic arrays, the so-called skyrmion crystal/lattice. Isolated skyrmion was observed in ferromagnetic $Fe_{0.5}Co_{0.5}Si$ films\cite{Yu}. A skyrmion is characterized by a topological charge, which imposes an energy barrier that protects it from decaying to the ground state. By virtue of their robust stability and small size, skyrmions have been faced as candidates for carrying information in the next generation of technological devices\cite{Fert}. Actually, skyrmions can be written, read, and deleted by using electric current \cite{Fert, Ezawa} passing through thin-film prototypes. However, the so-called skyrmion Hall effect (SkHE) bends their trajectories away from the driving current direction due to the Magnus force \cite{Cortes, Kim}, putting a severe obstacle on skyrmion usefulness.

A way to eliminate this undesirable effect is to depart to antiferromagnetic (AFM) systems \cite{Duine, Gobel}, where the Magnus force on each AFM sub-lattices offset. Recently, it was experimentally observed magnetic skyrmions in ferrimagnetic $GdFeCo$ films \cite{Woo} which have similar structures to antiferromagnets. It has been also claimed that a suitable spin-polarized current (SPC) could induce the appearance of AFM skyrmions\cite{Velkov} and that they could move for very long distances parallel to such a current\cite{Barker, Ricardo, Clodoaldo-skyrmion-JMMM}, making them feasible information carriers for technological skyrmion-based devices. In this sense, it was shown that skyrmions lying in AFM thin disks are promising candidates to a new generation microwave signal generators \cite{Shen}.  In this work, we show that an AFM skyrmion, emerging as the ground state in a thin small disk is driven by spin-polarized alternating current. Its dynamics is oscillatory with the same frequency as the applied current and the amplitude of the oscillations is kept unchanged while the current is on. We also analyze the action of a defect line inserted in the nanodisk. Defects have important effects on the topological objects\cite{Winder2008,Pereira2007}. Whenever we consider a heterogeneous disk, made from the junction of two half-disks with different exchange couplings, skyrmion dynamics remain essentially the same as before, but its core experiences a major effect: its core size and shape changes whenever trespassing the borderline, resembling a skyrmion ``breathing''.

\section{Model and Methods}

In order to describe the thin magnetic nanodisks, we consider a two-dimensional square lattice inside a circumference of radius $R$, in the \textit{xy}-plane. Our model Hamiltonian reads like below:
\begin{equation}\label{eq1}
	\mathcal{H} = \mathcal{H}_{\text{Exc}} + \mathcal{H}_{\text{DMI}} + \mathcal{H}_{\text{Ani}} + \mathcal{H}_{\text{Dip}},
\end{equation}
where
\begin{equation}
	\begin{split}
		\mathcal{H}_{Exc} &= +\sum_{m=1,2}A_{m} \sum_{\{i,j\}\in m} \vec{\mu}_{i} \cdot \vec{\mu}_{j} + A_{1-2} \sum_{i} \vec{\mu}_{i} \cdot \vec{\mu}_{i+1}\\
		\mathcal{H}_{\text{DMI}} &=  \sum\limits_{i,j}\vec{D}_{ij}\cdot\left(\vec{\mu}_{i}\times \vec{\mu}_{j}\right),\\
		\mathcal{H}_{\text{Ani}} &=
		-k_{u}\sum\limits_{i}\left(\mu_{i}^{z}\right)^{2},\\
		\mathcal{H}_{\text{Dip}} &= -d \sum\limits_{i,j}\left[\frac{3\,(\vec{\mu}_{i} \cdot \vec{r}_{i,j})\,(\vec{\mu}_{j} \cdot \vec{r}_{i,j})}{r_{i,j}^{5}}-\frac{\vec{\mu}_i \cdot \vec{\mu}_{j}}{r_{i,j}^{3}}\right]\nonumber
	\end{split}
\end{equation}
Here, $\vec{\mu}_{i}=\vec{M}_{i}/M_{s}=\mu_{i}^{x}\hat{x} + \mu_{i}^{y}\hat{y} + \mu_{i}^{z}\hat{z}$ is the magnetic moment unit vector at position $i$ ($M_{s}$ is the saturation magnetization of each AFM sub-lattice) while $m=1,2$ accounts for the two half-disks (mediums),  with exchange constants $A_1$ and $A_2$ (both of them positive to reinforce anti-alignment of neighbor spins), while the sum $\{i,j\}\in m$ is performed over nearest-neighbor spins of each half-disk $m=1,2$. Every spin (dipole) inside each medium has a coordination number of four, except those closest to the edges of the external half-circular boundary and those belonging to sites along the interface between the two mediums. In the latter case, each of these spins interacts with only three others of the same half-disk, while its remaining coupling is done with the nearest-neighbor of the other half-disk. This fact is accounted for by the mixing exchange parameter $A_{1-2}$-term, which only indexes spins belonging to the borderline. This is how our model describes a thin small circular disk composed of two distinct parts, each of them with its own exchange coupling.
\begin{figure}[H]
	\centering
	\includegraphics[width=80.0mm]{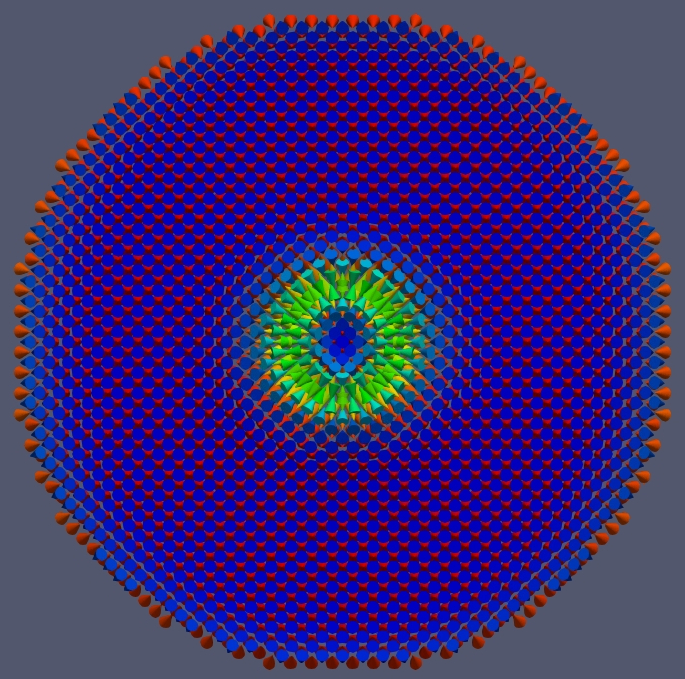}
	\caption{(Color online) A single AFM skyrmion emerges as the ground state in a small thin disk. Above, we have taken a homogenous disk, say $A_1=A_2=A_{1-2}=+1$, with diameter $65a$ and skyrmion circular core radius reads $\lambda=3a$. We have also taken $D_{i,j}=0.15$, $k_{u}=0.03$ and $d=0.01$.}
	\label{fig1}
\end{figure}
Clearly, if we set $A_1=A_2=A_{1-2}$, then we have a homogeneous disk with no interface. As an atomistic-type model we also adopt units as it is  usually done, for instance, $\hbar=c=1$ etc. However, for the sake of comparison, we quote SI units concerning our modeling. Typical AFM material exchange constants goes around $A_m\sim 10^1-10^2 {\rm meV}\sim 10^{-21} {\rm Joules}$. Frequency, $\omega$, is measured in units of $(A_m/\hbar) \sim 10^{12}-10^{13} {\rm Hz}$, while current density, $j_{0}$, is given in terms of {\em surface charge density} $\times$ {\em frequency}. For typical conductors one has $\sigma_e\sim 10^{-1}{\rm C/m^2}$, so that $(\sigma_e\omega)\sim 10^{11}-10^{12} {\rm A/m^2}$.

The term $\mathcal{H}_{\text{DMI}}$ accounts for the Dzyaloshinskii–Moriya interaction (DMI), where $\vec{D}_{ij}$ is the DMI vector. Here, we consider the N\'{e}el-type interface DMI interaction, say, $\vec{D}_{ij}$ is parallel to the plane of the system. The third term, $\mathcal{H}_{\text{Ani}}$,  represents uniaxial single-ion anisotropy, with easy-axis along $z$-direction, while $k_{u}$ is the anisotropy constant. The last contribution, $\mathcal{H}_{\text{Dip}}$, is the long-range dipole-dipole interaction, with constant parameter $d$.

Our numerical approach is split into two stages: Firstly, the stable AFM skyrmion configuration is obtained on the homogeneous disk at zero temperature by relaxation methods (Fig.\ref{fig1}). This is accomplished by using the skyrmion-like solution of the $O(3)$ nonlinear-$\sigma$
model \cite{Seibold}.

\begin{equation}
	\begin{split}
		\mu_{x} = (-1)^{i+j} \frac{\lambda i}{i^2 + j^2 + \lambda ^2},\\
		\mu_{y} = (-1)^{i+j} \frac{\lambda j}{i^2 + j^2 + \lambda ^2},\\
		\mu_{z} = (-1)^{i+j} \frac{1}{2} \frac{i^2 + j^2 - \lambda ^2}{i^2 + j^2 +\lambda ^2}
	\end{split}
\end{equation}
where $\lambda$ denotes the AFM skyrmion core size (radius). Figure \ref{fig1} shows such a configuration, with $\lambda=3a$, emerging as the ground state in a disk of diameter $65a$ ($a$ is the lattice spacing).

After having stabilized the skyrmion, the fourth-order Runge-Kutta method is employed to compute the dynamics of the magnetic moment, $\vec{\mu}_{i}$, using the Landau-Lifshitz-Gilbert equation,
\begin{equation}\label{eq2}
	\frac{\partial\vec{\mu}_{i}}{\partial t} = -\gamma\vec{\mu}_{i}\times
	\hat{H}_{\text{eff}}^{i}+\alpha \vec{\mu}_{i}\times\frac{
	\partial\vec{\mu}_{i}}{\partial t}
\end{equation}
where $\gamma$ is the gyromagnetic ratio, $\hat{H}_{\text{eff}}^{i}=-\frac{1}{\mu_{i}}\frac{\partial\mathcal{H}}{\partial \vec{\mu}_{i}}$ is the effective field on each spin, and $\alpha$ is the Gilbert damping coefficient. Spin-polarized current is introduced by using the Berger spin-transfer torque\cite{Brattas}:
\begin{equation}\label{eq3}
	\vec{\tau}_{B}= p\left(\vec{j}\cdot\nabla\right)\vec{\mu}\,,
\end{equation}
and
\begin{equation}\label{eq4}
	\vec{\tau}_{B\beta}= p\, \beta\, \vec{\mu}\times\left(\vec{j}\cdot\nabla\right)\vec{\mu}\,,
\end{equation}
which take into account the adiabatic and non-adiabatic torque respectively, $p$ is the spin polarization of the electric current density $\vec{j}$, while $\beta$-parameter characterizes its relative strength to the Berger torque, Eq.(\ref{eq3}). In our simulations, we have taken $\gamma = 1$, $\alpha=0.1$, $p = - 1$ and $\beta = 0$, since non-adiabatic torque is generally negligible compared to its adiabatic counterpart \cite{Nagaosa1,Nagaosa2}.

\section{Results and Discussion}

Firstly, we shall study the dynamics of the skyrmion emerging as the ground state in a homogeneous disk (Fig. \ref{fig1}). For that we have set $A_1=A_2=A_{1-2}=+1$ along with $D_{i,j}=0.15$, $k_{u}=0.03$ and $d=0.01$ (these parameters have been shown to be the optimal values yielding this skyrmion configuration in this disk size). Once the skyrmion is stabilized, spin-polarized alternating currents (SPACs) are applied along both disk directions, say, $\vec{j}_{x}=j_{0}\cos(\omega t)\hat{x}$ and $\vec{j}_{y}=j_{0}\sin(\omega t)\hat{y}$, with $j_{0}=0.11$ (in units of $\sigma_{e}\omega \sim 10^{11} - 10^{12}$ A/m$^2$, then $j_{0}\sim 1.1\times10^{10} - 10^{11}$ A/m$^2$, while $\omega=0.02 A_{1}/\hbar \sim 2\times 10^{10} - 10^{11}$ Hz). Here, we must emphasize that the current is homogeneous throughout the whole disk.

\begin{figure}[H]
	\centering
	\subfigure[$t=T/4$]{\includegraphics[width=40.0mm]{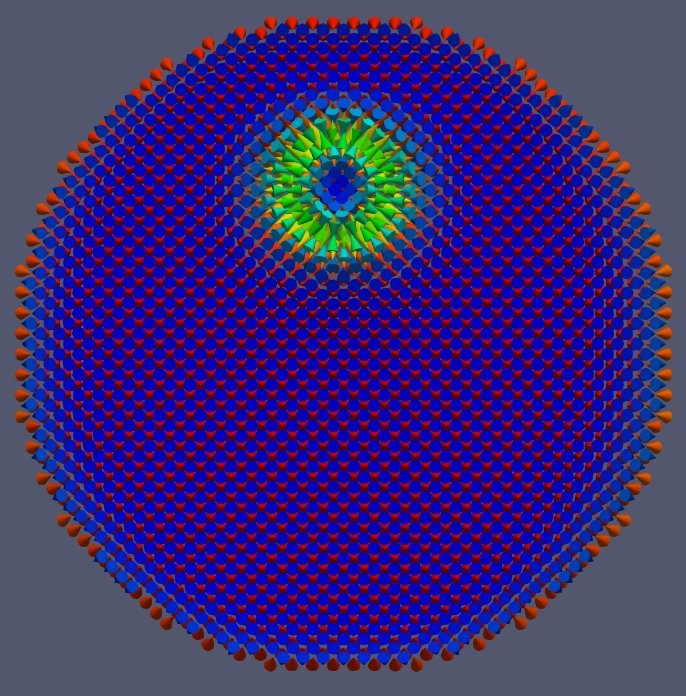}}
	\subfigure[$t=T/2$]{\includegraphics[width=40.0mm]{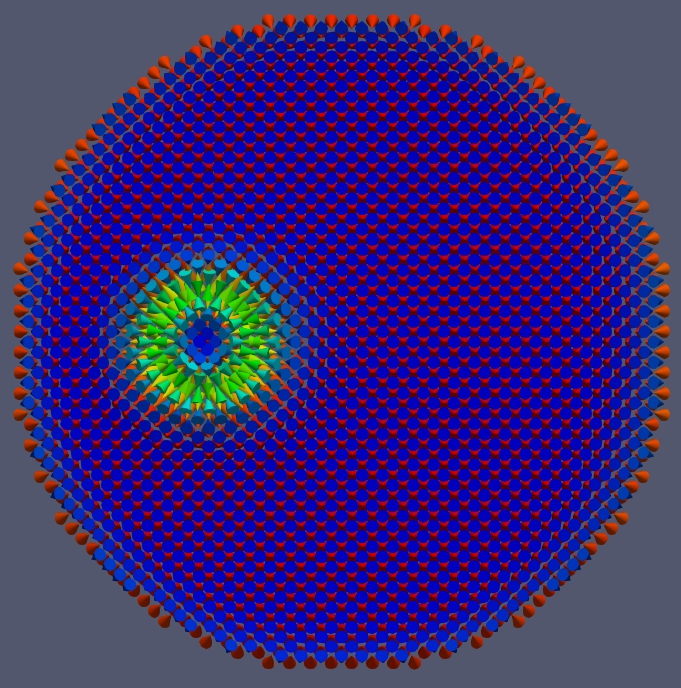}}
	\vskip 0.1cm
	\subfigure[$t=3T/4$]{\includegraphics[width=40.0mm]{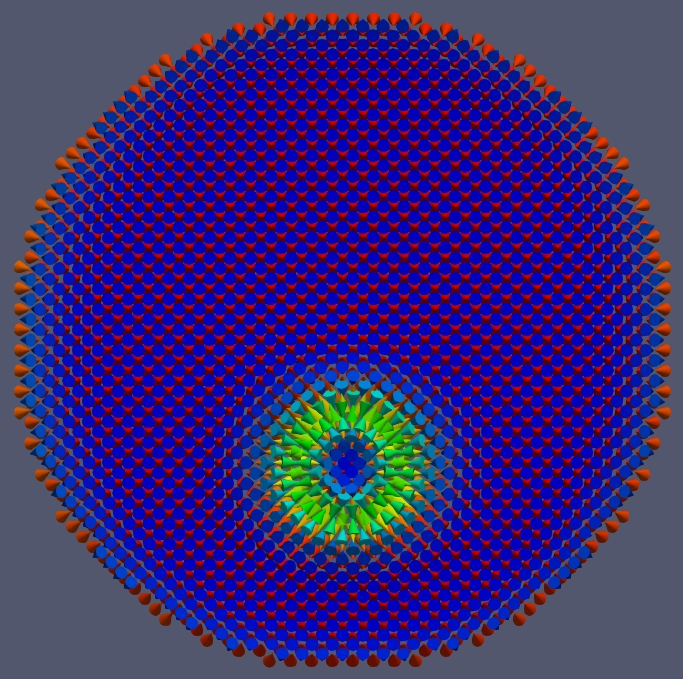}}
	\subfigure[$t=T$]{\includegraphics[width=40.0mm]{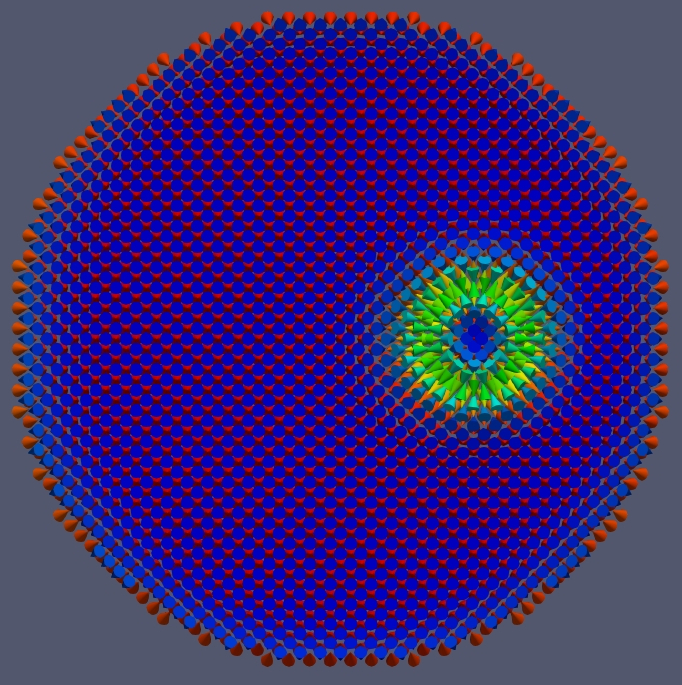}}
	\includegraphics[width=80mm]{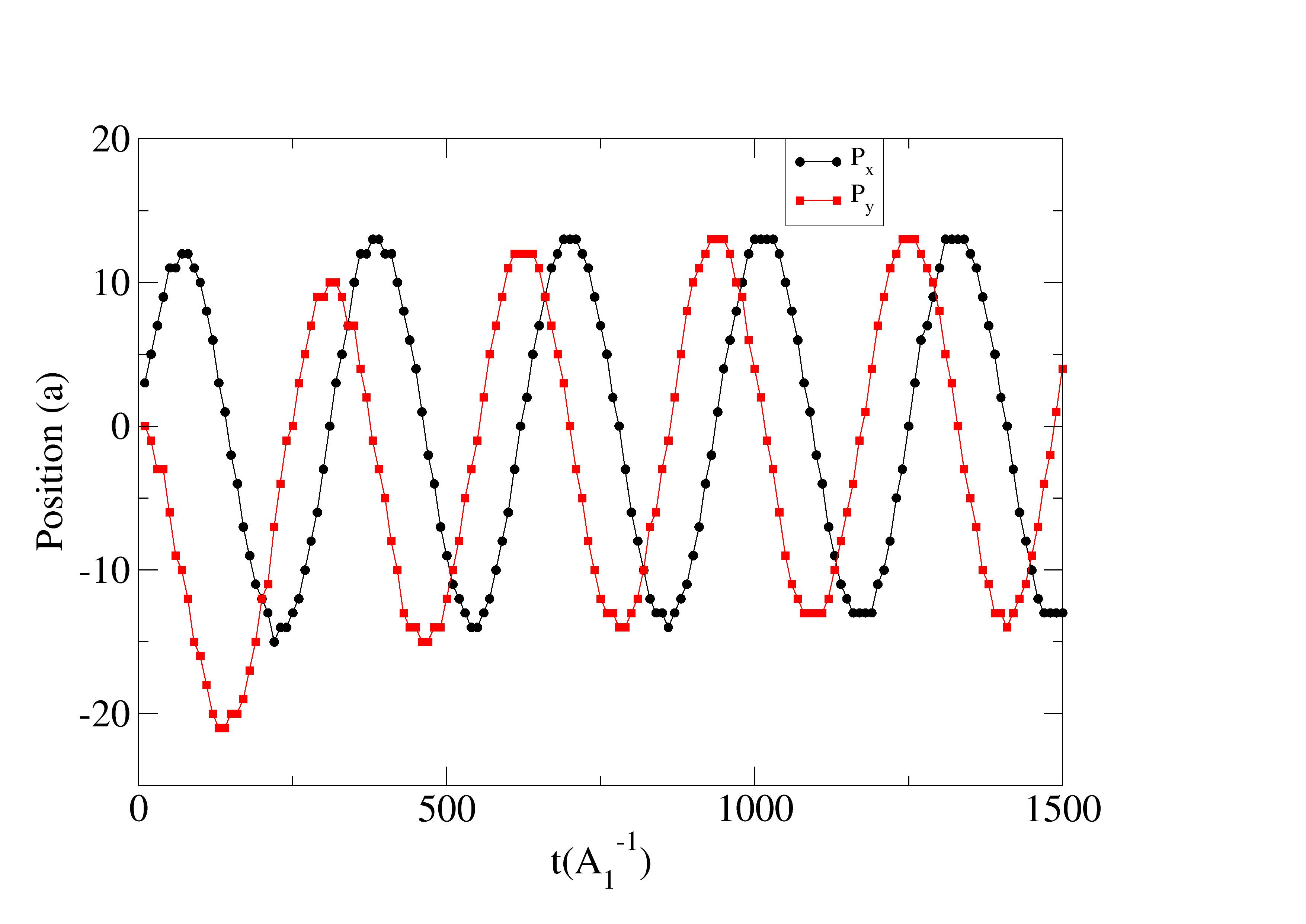}\\
	\caption{\label{skyrmion1} (Color online) The circular oscillatory dynamics of an AFM skyrmion driven by SPACs along $x$ and $y$ directions. Panels (a)-(d) show skyrmion position at different times ($T$ is the period of a complete cycle). At the bottom panel, its position is plotted as a function of time for several cycles. The skyrmion oscillates at the same frequency as the applied current, say, $\omega_{\rm sk}=\omega=0.02 A_{1}$, keeping amplitude practically unaltered. Such a motion is kept whenever SPAC is on, falling off immediately as the current is turned off.}
\end{figure}

Its torque immediately yields a circular oscillation of the AFM skyrmion, as shown in Fig. \ref{skyrmion1} (Movie 1, Supplementary Material, presents its complete dynamics in more detail). Whenever the current is turned off, skyrmion motion ceases abruptly and it moves towards the disk center, recovering its original static configuration. Note that the skyrmion oscillates under the SPAC action without any dissipation, once its amplitude remains practically unchanged over several oscillations, as illustrated in Fig.\ref{skyrmion1}. Moreover, SPAC provides controlled skyrmion oscillation: its amplitude and frequency may be adjusted on demand just by tuning the strength and frequency of the applied current, making our proposed system a feasible and practical way to control skyrmion dynamics by purely electric means. An analogous driven dynamics has been recently predicted for AFM-vortex pattern in thin small disks \cite{PLA2020-AFM-vortex-disk}.

\begin{figure}[H]
	\centering
	\includegraphics[width=80.0mm]{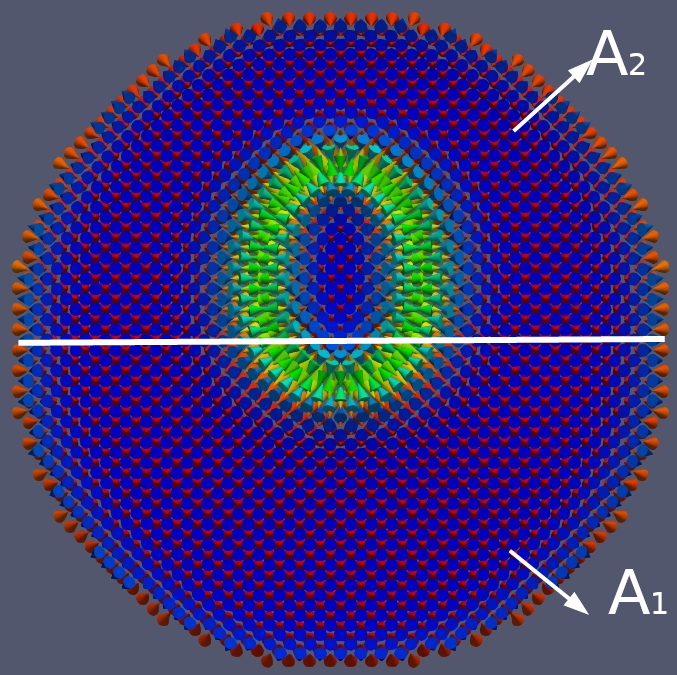}
	\caption{\label{skyrmion3} (Color online) An AFM skyrmion pattern in disk with $A_1=+1$ and $A_2=+0.8$, yielding a skyrmion as the ground state nucleated out of the disk center, at  ($x_0=0, y_0= 10a$), with core size (major elliptical axis) $\lambda \approx 6.2a$. This shift in the stable position and size increasing takes place as exchange cost decreases. [White line crossing the disk center along its horizontal direction, $y=0$, represents the interface].}
\end{figure}

Now, we depart to consider how an interface, along with two different half-disks, affects the skyrmion structure and dynamics. For that we shall keep the values for the other parameters as before,  say $D_{i,j}=0.15$, $k_{u}=0.03$,  and $d=0.01$. We also fix $A_1=+1$ and let $A_2$ and $A_{1-2}=(A_1+A_2)/2$ vary in $(0,1)$ interval. In our simulations, we have realized that the skyrmion tends to stabilize with its core region displaced to the medium with less exchange cost. Additionally, its core region is now elongated resembling an elliptical shape, see Fig. \ref{skyrmion3}. In addition, we have also noticed that $A_{2-cr}= +0.75$ comes to be a critical value, namely, associated to the skyrmion dynamics and its capability of crossing the interface. Indeed, using the same SPAC frequency and magnitude as before, $\omega= 0.02$ and $j_0=0.11$, we have observed that for $A_2>0.75$ the skyrmion crosses the interface and reaches to $A_1$-medium, whereas if $A_2<0.75$ the energy coming from this specific current is not sufficient to promote its crossing, and skyrmion is trapped in the half-disk with lower exchange constant.

\begin{figure}[H]
	\centering
	\subfigure[$t=T/4$]{\includegraphics[width=40.0mm]{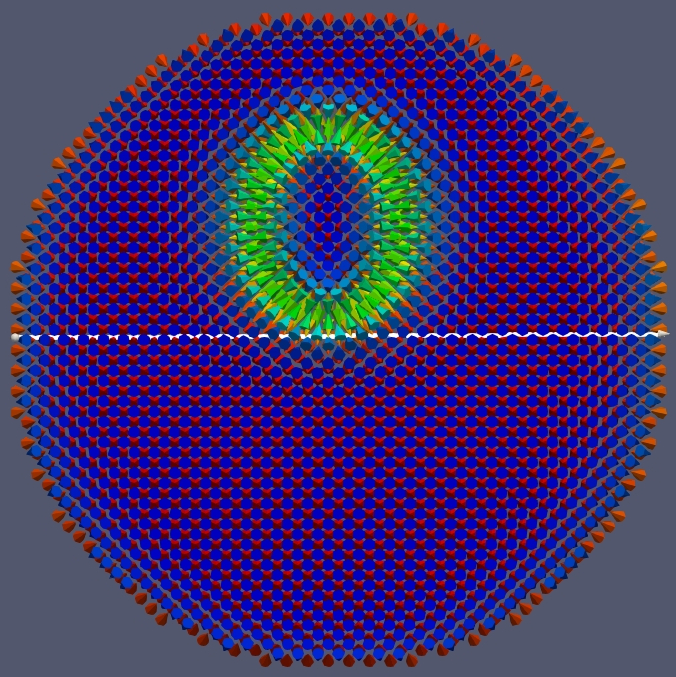}}
	\subfigure[$t=T/2$]{\includegraphics[width=40.5mm]{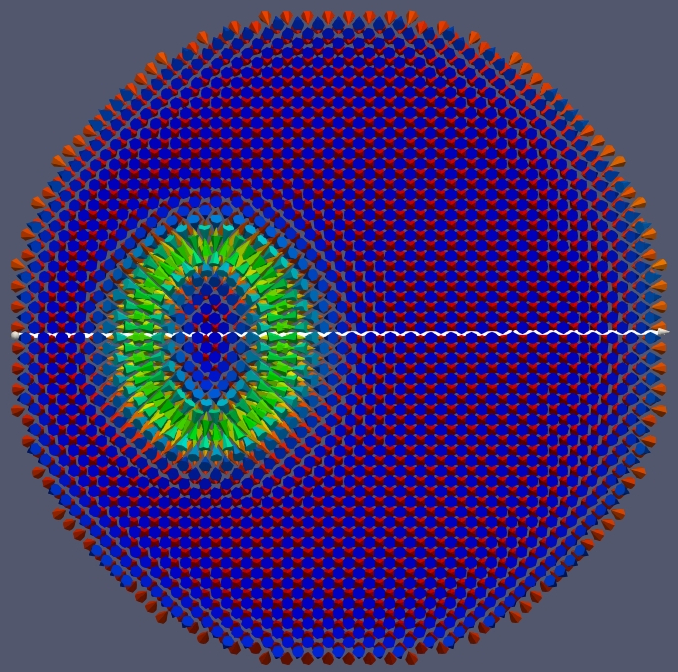}}
	\vskip 0.1cm
	\subfigure[$t=3T/4$]{\includegraphics[width=40.0mm]{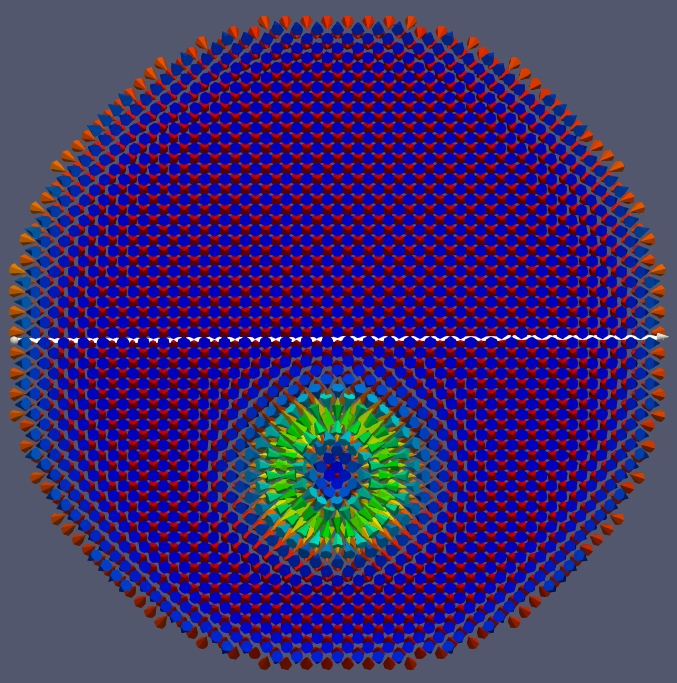}}
	\subfigure[$t=T$]{\includegraphics[width=40.5mm]{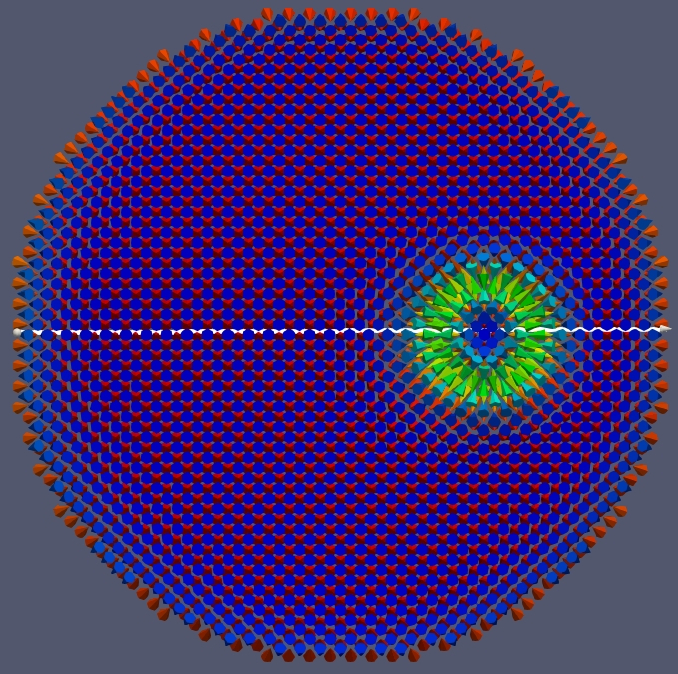}}
	\vskip .2cm
	\includegraphics[width=80mm]{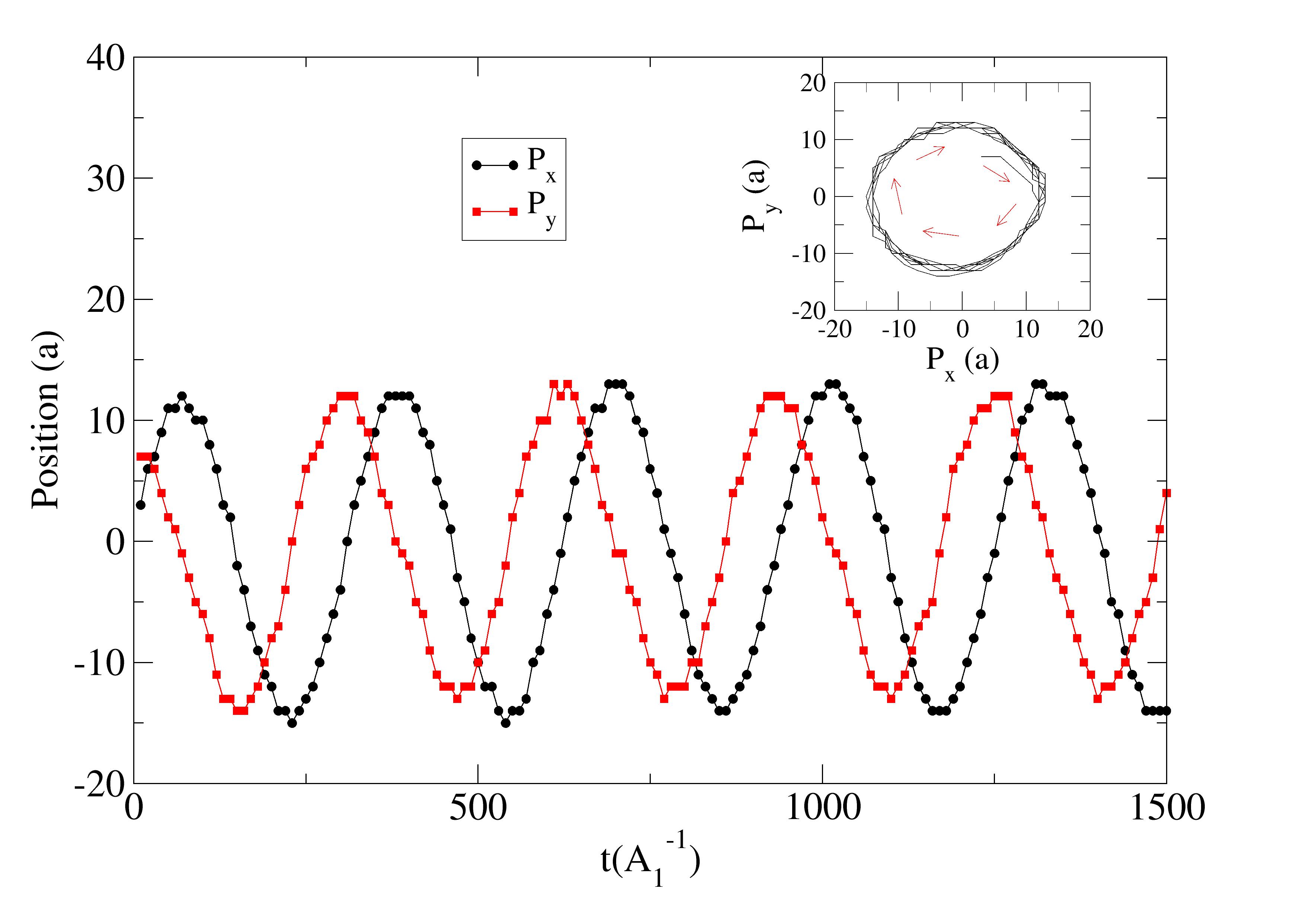}\\
	\caption{\label{skyrmion4} (Color online) Skyrmion oscillation in a heterogeneous disk. Panels (a)-(d) show skyrmion position at different times ($T$ is the period of a complete cycle). At the bottom panel, its position is plotted against time. As occur in the homogeneous disk, skyrmion oscillates with applied current frequency, $\omega_{\rm sk}=\omega=0.02 A_{1}$, keeping amplitude practically unaltered, without energy dissipation. However, now the skyrmion trajectory is no longer perfectly circular, it rather resembles an ellipse (with small eccentricity, see inset; note the slight difference between its amplitude along $x$ and $y$ directions over the cycles). Once current is turned off skyrmion motion quickly falls off.}
\end{figure}

Actually, whenever skyrmion energy is enough to cross the interface barrier, its dynamics is performed practically without dissipation. Indeed, we have realized that due to its interaction with the interface, its core deforms absorbing (delivering) energy from (to) the remaining portions of the skyrmion keeping its dynamical stability as a whole. [Recall that each half-disk has a different exchange strength, so that one 'half-skyrmion' costs higher than the other lying on the other side of the interface]. More specifically, when it crosses from $A_2$ to $A_1$ medium, its spins rearrange to shorten its core size, while whenever crossing back the core is enlarged as a result of the relaxation of the spins. As a whole, the crossing processes imply in a sort of {\em breathing} of the AFM skyrmion with remarkable effects on its core region. Some snapshots of such a skyrmion behavior is shown in Fig.\ref{skyrmion4}, whereas its complete oscillatory motion is presented in  Movie $2$ (Supplementary Material). For the sake of completeness, we have simulated lower $A_2$-values with this disk diameter, $65a$ (all other energetic constants have kept unaltered, namely $A_1=+1$). For $0.7\le A_2\le 0.73$ a current strength $j_0\approx 0.22$ (doubled the former one) along with a larger frequency $\omega\approx 0.03$ yield driven oscillation like aforementioned. On the other hand, whenever $A_2\leq 0.69$ we have not obtained any current protocol providing skyrmion oscillation. In this case, lower current parameters are not sufficient to overcome the trapping brought about by exchange cost difference (as discussed above), whereas high enough ones yield skyrmion displacing to the circular disk border, where it cannot keep its topological charge and is annihilated. The main lesson is: lower $A_2$-values demands larger disks so that higher SPAC parameters have enough space to drive the skyrmion dynamics in a controlled way.

\begin{figure}[H]
	\begin{center}
	\includegraphics[width=100.0mm]{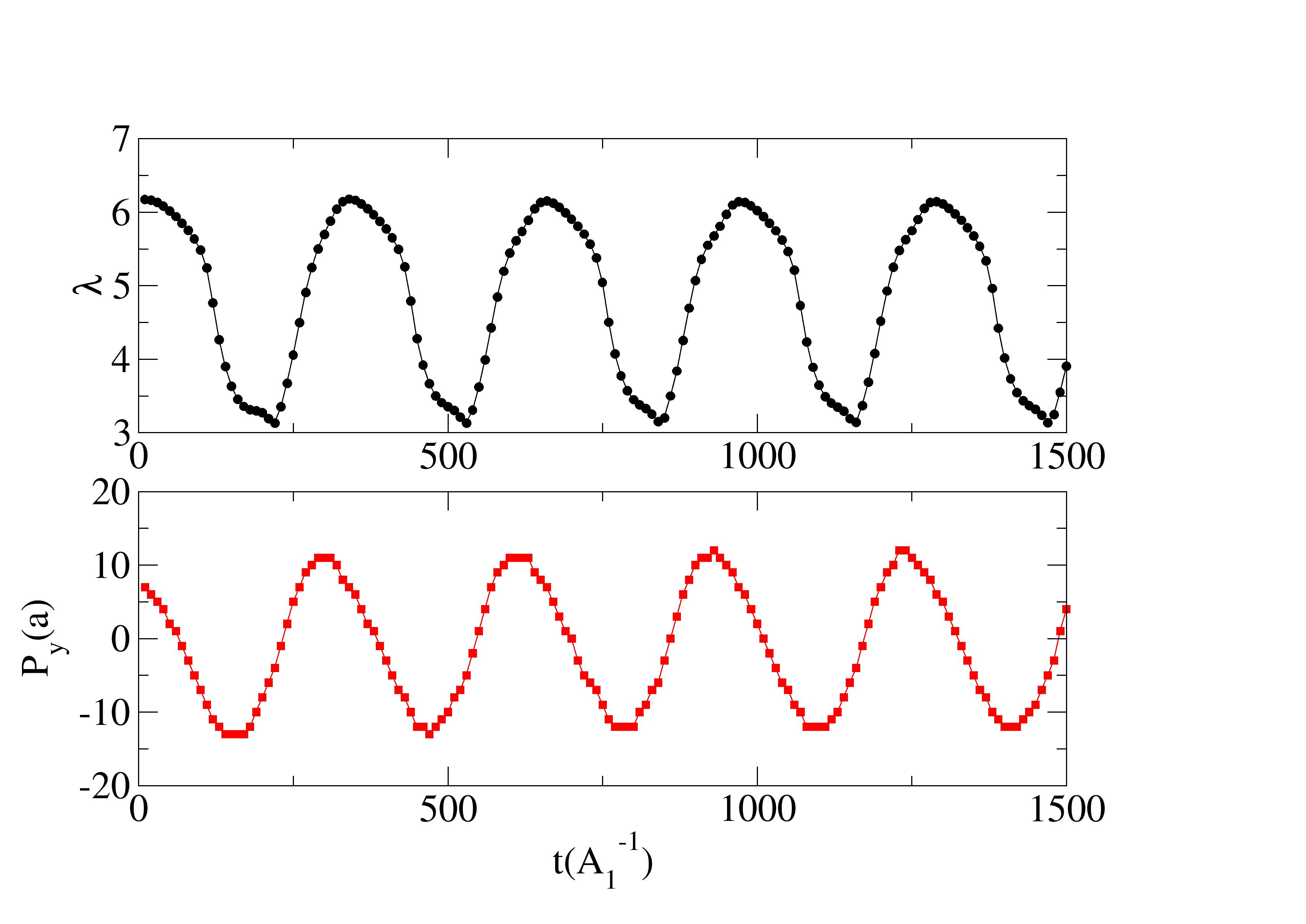}
	\caption{\label{fig2} How the skyrmion size (major axis length), $\lambda$, and position along $y$, $P_{y}$, oscillate with time. Whenever the skyrmion core is located at the upper half-disk its core region is enlarged in comparison with the size at higher exchange cost, lower half-disk. Note that for a short time interval around interface crossing skyrmion core size experiences a considerable variation.}
	\end{center}
\end{figure}

\section{Concluding remarks}

We have shown that an AFM skyrmion confined to the geometry of a thin small disk may be driven by an alternating spin-polarized current which puts it to oscillate at current frequency. Its oscillation takes place practically without dissipation, keeping the amplitude as long as the current is switched on. Whenever a borderline is inserted separating two half-disks with distinct exchange constants, we have two main situations depending on the ratio between them. For a disk $65a$-diameter and SPAC parameters $(j_0=0.11;\, \omega=0.02)$, if $A_2/A_1<0.75$, the skyrmion is trapped in the $A_2$ half-disk and it does not move at all. Nevertheless, if $0.75<A_2/A_1<1$ the interface does not jeopardize skyrmion oscillation at all. Indeed, interface mainly affects skyrmion core changing its profile whenever crossing the interface. We claim that such a robustness of AFM skyrmion dynamics driven by spin-polarized current may be useful for its potential application in the framework of the emerging branch of topological antiferromagnetism.

\section{Acknowledgements}

The authors thank CAPES (financial code 001), CNPq, and FAPEMIG (Brazilian agencies) for partial financial support.


\end{document}